\documentclass[seceq]{ptptex}

\usepackage{graphicx}
\usepackage{wrapft}

\newcommand{\sol}{M_{\odot}}

\newcommand{\rscript}[1]{\mbox{\scriptsize #1}}
\newcommand{\rhoh}{\rho_{_H}}
\newcommand{\rhob}{\rho_{_B}}
\newcommand{\rhon}{\rho_{_N}}

\newcommand{\gammaT}{\widetilde{\gamma}}
\newcommand{\GammaT}{\widetilde{\Gamma}}

\newcommand{\KT}{\widetilde{K}}
\newcommand{\KH}{\widehat{K}}
\newcommand{\AT}{\widetilde{A}}
\newcommand{\JT}{\widetilde{J}}
\newcommand{\DT}{\widetilde{D}}
\newcommand{\RT}{\widetilde{R}}
\newcommand{\LaplaceT}{\widetilde{\triangle}}
\newcommand{\auz}{\alpha u^0}
\newcommand{\sgamma}{\sqrt{\gamma}}
\newcommand{\FT}{\widetilde{F}}

\newcommand{\Dtv}[2]{\partial_t #2 #1^{l} \partial_l #2}

\newcommand{\STF}[1]{#1^{\mbox{\scriptsize STF}}}
\newcommand{\St}{\sin \!\theta \,}
\newcommand{\Sts}{\sin^2 \!\theta \,}
\newcommand{\Ct}{\cos \theta \,}
\newcommand{\Cts}{\cos^2 \! \theta \,}
\newcommand{\Sp}{\sin \! \phi \,}
\newcommand{\Sps}{\sin^2 \! \phi \,}
\newcommand{\Cp}{\cos \phi \,}
\newcommand{\Cps}{\cos^2 \! \phi \,}
\newcommand{\dOmega}{\mbox{d}\Omega}

\notypesetlogo                       

\title{%
General Relativistic Numerical Simulation on Coalescing Binary Neutron
Stars and Gauge-Invariant Gravitational Wave Extraction%
}

\author{
Mari \textsc{Kawamura}$^{1}$
Ken-ichi \textsc{Oohara}$^{2}$
and
Takashi \textsc{Nakamura}$^{3}$

}

\inst{%
$^1$ Graduate School of Science and Technology, Niigata University,
  Niigata, 950--2181, Japan\\
$^2$ Department of Physics, Niigata University,
         Niigata 950--2181, Japan\\
$^3$ Department of Physics, Kyoto University,
         Kyoto 606-8502, Japan
}

\abst{%
  We are developing 3 dimensional simulation codes for coalescing
  binary neutron stars. A  code using the maximal slicing condition is
  obtained. To evaluate the gravitational radiation, we implemented a
  gauge-invariant wave extraction and compared the wave forms with the
  metric tensors at the wave zone. The energy spectrum of the waves
  was also evaluated to investigate the possibility that the
  excitation of the quasi-normal modes of the black hole, which may be
  formed after the merger of two stars, can be observed.}

\begin{document}

\maketitle

\section{Introduction}

We are constructing computer codes on 3D numerical
relativity.\cite{ONS97,ON99} We report here on the present status of
our computer code development to study the fully general relativistic
evolution of spacetimes and matter. Our main target is to study the
evolution of coalescing binary neutron stars and the radiation of
gravitational waves from the merger. Coalescing neutron stars or black
hole binaries are the most promising sources of strong gravitational
waves for interferometric detectors such as TAMA300, LIGO, VIRGO, and
GEO600.\cite{DETECT} The first detection by these detectors may be the
waves from a black hole binary because of the larger mass of the
system. However the astrophysical importance of a coalescing neutron
star binary is not smaller than a black hole binary. For example, in a
certain model of a gamma ray bursts\cite{HPA91,RM92}, the central
engine is a coalescing neutron star binary. Furthermore, a detailed
analysis of the detected waves near the merger of two stars will give
information on the size of a neutron star and then on the equation of
state of high density matter.

Gravitational waves from coalescing binaries consist of three phases;
(1) the inspiral phase, (2) the merging phase and (3) the ringing-down
phase.  At the inspiral phase, the separation between two stars is so
large that they may be considered as point masses and the luminosity
of gravitational waves is so small that the orbit of each star may be
quasi-stationary. The waves from this phase is called the chirp signal
and the wave form can be estimated in a post-Newtonian approximation.
The coalescing binary neutron stars at the merging and the
ringing-down phase cannot be considered as point masses and the
general relativistic effects become large. Thus general relativistic
numerical simulations are necessary to investigate the evolution of
the binary at these phases and to predict the radiation of
gravitational waves from the merger.

In the previous code we used the conformal slicing condition in which
the metric becomes the Schwarzschild one in the outer vacuum region so
that the wave extraction is easy, while for the long time integration
the slicing is not stable.\cite{ONS97} Then we started to construct a
new code using the maximal slicing condition. Disadvantages of the
maximal slicing compared with the conformal slicing are 1) the
extraction of gravitational waves is not straightforward, 2) more CPU
hours are required since an elliptic partial differential equation
should be solved for the lapse function $\alpha$.  For the former,
Abrahams et.al.\cite{ADHS92} gave the gauge-invariant wave extraction
technique for axially symmetric, even-parity perturbations in the
Schwarzschild metric. We extend their technique to non-axially
symmetric perturbations both for even- and odd-parities. As a result,
we found that the gauge dependent modes in a spatial metric
perturbation is so small at the vacuum exterior region that a good
estimate of gravitational waves is given easily. Shibata and
Uryu\cite{SU02} have reported gravitational waves from the merger of
binary neutron stars using a different code with different coordinate
conditions so that the comparison of our results with theirs is
important.

In the followings, we show recent results of our 3D general
relativistic simulations of coalescence of binary neutron stars and
the estimation of gravitational waves. In \S 2 and \S 3, we describe
basic equations for our general relativistic code for coalescing
binary neutron stars and our coordinate conditions. Numerical methods
are described in \S 4 and we show how to extract gauge-invariant
gravitational wave form in \S 5. In \S 6, numerical results of
simulations of the coalescing neutron star binary are presented and \S
7 is devoted to the discussions.

\section{Basic Equations}

In order to follow the time evolution of the space-time and the
matter, the Einstein equation should be reduced to a system of
evolution equations. We use the (3+1)-formalism of the Einstein
equation.  In the (3+1)-formalism of the Einstein equation, the line
element is given by
\begin{equation}
  ds^2 = - \alpha^2 dt^2 
  + \gamma_{ij} ( dx^i + \beta^i dt ) ( dx^j +  \beta^j dt),
\end{equation}
where $\alpha$, $\beta^i$ and $\gamma_{ij}$ are the lapse function,
the shift vector and the intrinsic metric of 3-space, respectively.
The Einstein equation is decomposed to the four constraint equations
and 12 evolution equations. The Hamiltonian and the momentum
constraint equations are
\begin{equation}
  R + K^2 - K_{ij} K^{ij} = 16 \pi \rhoh ,
  \label{eq:hconst}
\end{equation}
and
\begin{equation}
  D_j \left( K^j{}_i - \delta^j{}_i K \right) = 8 \pi J_i ,
  \label{eq:mconst}
\end{equation}
respectively. The evolution equations are
\begin{equation}
  \label{eq:evolgij}
  \partial_t \gamma_{ij} = - 2\alpha K_{ij} + D_i \beta_j + D_j
  \beta_i 
\end{equation}
and
\begin{eqnarray}
  \partial_t K_{ij} & = & \alpha \left[ R_{ij} - 8\pi \left\{ S_{ij}
      + \mbox{$\frac{1}{2}$} \gamma_{ij} \left( \rhoh - S^{l}{}_{l}
      \right) \right\} \right] - D_i D_j \alpha \nonumber \\
  & & + \alpha \left( K K_{ij} - 2 K_{il} K^{l}{}_j \right)
  + K_{l i} D_j \beta^l + K_{l j} D_i \beta^l + \beta^l D_l K_{ij} ,
  \label{eq:evolkij}
\end{eqnarray}
where $R_{ij}$ , $R = \gamma^{ij}R_{ij}$, $K_{ij}$ and $K =
\gamma^{ij} K_{ij}$ are the 3-D Ricci tensor, its trace, the extrinsic
curvature tensor, its trace, respectively and $D_i$ is the covariant
derivative associated with $\gamma_{ij}$. The quantities $\rhoh$,
$J_i$ and $S_{ij}$ are the energy density, the momentum density and
the stress tensor, respectively, measured by the observer moving along
the line normal to the spacelike hypersurface of $t =$ constant.

In order to give initial data, the constraint equations are solved for
given $\rhoh$ and $J_i$.  Here we assume that $K=0$ and $\gamma_{ij}$
is conformally flat at $t=0$ as
\begin{equation}
  \label{eq:cflat}
  \gamma_{ij} = \phi^4 \, \gammaT_{ij} ,
\end{equation}
where $\gammaT_{ij}$ is the flat space metric. Defining the conformal
transformation as
\begin{equation}
  \label{eq:conf-i}
    \KT_{ij} \equiv \phi^2 K_{ij} , \quad \KT_i{}^j \equiv \phi^6
    K_i{}^j , \quad \KT^{ij} \equiv \phi^{10} K^{ij} , \quad \rhob
    \equiv \phi^6 \rhoh , \quad \JT_i \equiv \phi^6
    J_i ,
\end{equation}
we can reduce Eq.(\ref{eq:mconst}) to
\begin{equation}
  \label{eq:mconst2}
  \DT_j \KT^j{}_i = 8 \pi \JT_i ,
\end{equation}
where $\DT_i$ is the covariant derivative associated with
$\gammaT_{ij}$. The traceless extrinsic curvature can be expressed as
the sum of the transverse traceless part $\KT^{\rscript{TT}}_{ij}$ and
the longitudinal traceless part as $(LW)_{ij}$;
\begin{equation}
  \KT_{ij} = \KT^{\rscript{TT}}_{ij} + (LW)_{ij} ,
\end{equation}
where
\begin{equation}
  (LW)_{ij} = \DT_i W_j + \DT_j W_i
  - \frac{2}{3} \, \gammaT_{ij} \DT^{l} W_{l} .
\end{equation}
Assuming $\KT^{\rscript{TT}}_{ij} = 0$, we have the momentum
constraint equations (Eq.(\ref{eq:mconst2})) as
\begin{equation}
  \label{eq:mconst3}
  \LaplaceT W_i + \frac{1}{3} \DT_i \DT^j W_j = 8 \pi \JT_i ,
\end{equation}
where $\widetilde{\triangle} \equiv \DT^i \DT_i$.

The Hamiltonian constraint equation (Eq.(\ref{eq:hconst})) becomes
\begin{equation}
  \widetilde{\Delta}\phi = - 2\pi\phi^{-1}\rho_{B} -
  \frac{1}{8}\phi^{-7}\KT_{ij}\KT^{ij}.
\label{eq:hconst2}
\end{equation}

To solve the evolution of the metric tensor, we define the following
variables as
\begin{equation}
  \label{eq:def-phi}
   \phi = \left( \mbox{det} (\gamma_{ij}) \right)^{\frac{1}{12}},
\end{equation}
\begin{equation}
  \label{eq:def-gammaTij}
   \gammaT_{ij} = \phi^{-4} \gamma_{ij} \ ,
\end{equation}
\begin{equation}
  \label{eq:def-GammaTi}
   \FT^{i} = \gammaT^{ij}{}_{,j}
\end{equation}
\begin{equation}
  \label{eq:def-KHij}
   \KH_{ij} = \phi^{-4} \STF{(K_{ij})},
\end{equation}
\begin{equation}
  \label{eq:def-trK}
  K = \gamma^{ij} K_{ij}
\end{equation}
where
\begin{equation}
   \gammaT^{ik} \gammaT_{kj} = \delta^i{}_j , \qquad
\end{equation}
and
\begin{equation}
  \label{eq:STF}
    \STF{(K_{ij})} \equiv \frac{1}{2} \left( K_{ij} + K_{ji}
      - \frac{2}{3} \gammaT_{ij} \gammaT^{kl} K_{kl} \right) .
\end{equation}
The indices of $\KH_{ij}$ are lowered and raised by $\gammaT_{ij}$ and
$\gammaT^{ij}$;
\begin{equation}
   \KH^i{}_j = \gammaT^{ik} \KH_{kj} \ ,
   \qquad \KH^{ij} = \gammaT^{jk} \KH^i{}_k .
\end{equation}
We treat $\phi$, $\gammaT_{ij}$, $\FT^i$, $\KH_{ij}$ and $K$ as
independent variables in numerical calculations, although they are not
really independent. In this framework, Eq.(\ref{eq:def-GammaTi}) and
the following equations can be treated as additional constraints,
\begin{equation}
  \label{eq:cons-a1}
  \mbox{det} (\gammaT_{ij}) = 1,
\end{equation}
\begin{equation}
  \label{eq:cons-a2}
  \gammaT^{ik} \KH_{ki} = 0.
\end{equation}

Recently some kinds of the reformulation of the Einstein equations in
numerical relativity have been proposed to obtain numerically stable
codes.\cite{SY03} Our formulation is the simplest one which initiated
such researches of reformulations. The motivation to use the
formulation is that we encountered numerical instability in
development of a 3-dimensional, fully relativistic numerical
simulation code and suppose that numerical errors in the second
derivative of the metric tensor needed to calculate the Ricci tensor
are likely to cause large errors and the instability. Then we decided
to compute $\FT^{i}$ as independent variables and calculate the Ricci
tensor using $\FT^{i}$ (see Eqs.(\ref{eq:RT1})--(\ref{eq:RT3})). Then
we found that such modification makes the code dramatically
stable. This formulation is now often cited as BSSN
formulation,\cite{SN95,BS99} but it was first introduced by Nakamura,
Oohara and Kojima,\cite{NOK87} and is continuously used in our code
development.\cite{ON99,ONS97,ON97,NO99}

From Eqs.(\ref{eq:evolgij}) and (\ref{eq:evolkij}), we have the
evolution equations for these variables:
\begin{equation}
  \label{eq:evolphi}
  \partial_t \phi =
  - \frac{\phi}{6} \left( \alpha K - D_l \beta^l \right) ,
\end{equation}
\begin{equation}
  \label{eq:evolgTij}
  \partial_t \gammaT_{ij} = - 2 \left[ \alpha \KH_{ij}
    - \phi^{-4} \STF{(D_i \beta_j)} \right] \equiv \AT_{ij} ,
\end{equation}
\begin{eqnarray}
  \partial_t \KH_{ij} & = & \phi^{-4}
  \left\{ \alpha \left[ \STF{(R_{ij})} - 8 \pi \STF{(S_{ij})} \right]
    - \STF{(D_i D_j \alpha)} \right\} \nonumber \\
  & + & \alpha \left( K \KH_{ij} - 2 \KH_{il} \KH^l{}_j \right)
     \label{eq:evolkHij} \\
  & + & \KH_{il} D_j \beta^l + \KH_{jl} D_i \beta^l + \phi^{-4}
  \beta^l D_l \left( \phi^4 \KH_{ij} \right) - \frac{2}{3} \KH_{ij}
  D_l \beta^l \nonumber
\end{eqnarray}
and
\begin{equation}
  \partial_t K = \alpha \! \left[ \KH_{ij} \KH^{ij} \! + \!
  \frac{1}{3} K^2 \! + \! 4\pi \! \left( \rhoh \! + \! S^i{}_i \right)
  \right] \! - \! D^i D_i \alpha + \beta^l D_l K. \label{eq:evoltrK}
\end{equation}
The quantity $\FT^{i}$ obeys
\begin{equation}
  \label{eq:evolGTi}
  \partial_t \FT^{i} = - \AT^{ij}{}_{,j} ,
\end{equation}
since
\begin{equation}
  \label{eq:ATuij}
  \partial_t \gammaT^{ij} = - \gammaT^{ik} \gammaT^{jl} \partial_t
  \gammaT_{kl}
  = - \gammaT^{ik} \gammaT^{jl} \AT_{kl} \equiv - \AT^{ij} .
\end{equation}

We assume that the matter is the perfect fluid, whose stress-energy is
given by
\begin{equation}
  \label{eq:enmom}
  T_{\mu \nu} = ( \rho + \rho \varepsilon + p ) u_{\mu} u_{\nu}
  + p g_{\mu \nu},
\end{equation}
where $\rho$, $\varepsilon$ and $p$ are the proper mass density, the
specific internal energy and the pressure, respectively, and $u_\mu$
is the four-velocity of the fluid. The energy density $\rhoh$, the
momentum density $J_i$ and the stress tensor $S_{ij}$ of the matter
measured by the normal line observer are, respectively, given by
\begin{equation}
  \label{eq:qhydro}
  \rhoh \equiv n^\mu n^\nu T_{\mu \nu}, \qquad
  J_i \equiv - h_i{}^\mu n^\nu T_{\mu \nu}, \qquad
  S_{ij} \equiv h_i{}^\mu h_j{}^\nu T_{\mu \nu},
\end{equation}
where $n_\mu$ is the unit timelike four-vector normal to the spacelike
hypersurface and $h_{\mu \nu}$ is the projection tensor into the
hypersurface defined by
\begin{equation}
  h_{\mu \nu} = g_{\mu \nu} + n_\mu n_\nu .
\end{equation}
The relativistic hydrodynamics equations are obtained from the
conservation of baryon number, $\nabla_\mu ( \rho u^\mu ) = 0$, and
the energy-momentum conservation law, $\nabla_\nu T_\mu{}^\nu{} =
0$. In order to obtain equations similar to the Newtonian
hydrodynamics equations, we define $\rhon$ and $u_i^N$ by
\begin{equation}
  \label{eq:rhonun}
  \rhon \equiv \sgamma \auz \rho , \qquad
  u_i^N = \frac{J_i}{\auz \rho},
\end{equation}
respectively, where $\gamma = \mbox{det} (\gamma_{ij})$. Then the
equation for the conservation of baryon number takes the form
\begin{equation}
  \label{eq:hydrob}
  \partial_t \rhon + \partial_l \left( \rhon V^l \right) = 0 ,
\end{equation}
where
\begin{equation}
  V^l = \frac{u^l}{u^0} = \frac{\alpha J^l}{p + \rhoh} - \beta^l .
\end{equation}
The equation for the momentum conservation is
\begin{eqnarray}
  \partial_t (\rhon u_i^N) + \partial_l \left( \rhon u_i^N V^l \right)
  & = & - \sgamma \alpha \partial_i p - \sgamma ( p + \rhoh )
  \partial_i \alpha \nonumber \\
  & & + \frac{\sgamma \alpha J^k J^l}{2(p + \rhoh)}
  \partial_i \gamma_{kl} + \sgamma J_l \partial_i \beta^l . 
  \label{eq:hydrom}
\end{eqnarray}
The equation for the internal energy conservation becomes
\begin{equation}
  \label{eq:hydroe}
  \partial_t (\rhon \varepsilon) + \partial_l \left( \rhon \varepsilon
  V^l \right) = - p \partial_\nu \left( \sgamma \alpha u^\nu \right).
\end{equation}
To complete the hydrodynamics equations, we need an equation of state,
\begin{equation}
  \label{eq:eos}
  p = p(\varepsilon, \rho).
\end{equation}

The right-hand side of Eq.(\ref{eq:hydroe}) includes the time
derivative. For a polytropic equation of state, $p = (\Gamma - 1) \rho
\varepsilon$, however, the equation reduces to
\begin{equation}
  \label{eq:hydroe2}
    \partial_t (\rhon \varepsilon_{_N})
    + \partial_l \left( \rhon \varepsilon_{_N} V^l \right)
    = - p_{_N} \partial_l V^l ,
\end{equation}
where
\begin{eqnarray}
  \label{eq:defen}
  \varepsilon_{_N} & = & \left( \sgamma \auz \right)^{\Gamma -1} \,
  \varepsilon , \\
  \label{eq:defpn}
  p_{_N} & = & (\Gamma - 1) \rhon \varepsilon_{_N}
  = \left( \sgamma \auz \right)^\Gamma p .
\end{eqnarray}

\section{Coordinate conditions}
\label{sec:CC}

The choice of the shift vector $\beta^i$ and the lapse function
$\alpha$ is important because the stability of the code largely
depends on them and because it is intimately related to the extraction
of physically relevant information, including gravitational radiation,
in numerical relativity.  Since the right-hand side of
Eq.(\ref{eq:evolgTij}), $\AT_{ij}$ is trace-free, the determinant of
$\gammaT_{ij}$ is preserved in time and the condition
\begin{equation}
  \label{eq:mind}
  D_j \AT^{ij} = 0
\end{equation}
produces the minimal distortion shift vector. This is a good choice of
the spatial coordinate, but it is too complicated to be solved
numerically. Instead we replace the covariant derivative in Eq. (3.1)
by the partial derivative as
\begin{equation}
  \label{eq:pmind}
   \partial_j \AT^{ij} = 0 .
\end{equation} 
In this condition, we can simply set
\begin{equation}
  \label{eq:GammaT0}
  \FT^i = 0,
\end{equation}
since Eq.(\ref{eq:evolGTi}) becomes $\partial_t \FT^i = 0$ and
$\FT^i(t = 0) = 0$.

From Eqs.(\ref{eq:evolgTij}), (\ref{eq:ATuij}) and (\ref{eq:GammaT0}),
Eq.(\ref{eq:pmind}) is reduced to
\begin{equation}
  \label{eq:ATuijj}
  \partial_j \AT^{ij} = -2 \partial_j \left( \alpha \KH^{ij} \right) +
  \gammaT^{jk} \partial_j \partial_k \beta^i + \frac{1}{3}
  \gammaT^{ij} \partial_j \partial_k \beta^k = 0,
\end{equation}
which yields an elliptic equation for the shift vector $\beta^i$:
\begin{equation}
  \label{eq:beta}
  \nabla^2 \beta^i + \frac{1}{3} \partial_i \partial_l \beta^l = 2
  \partial_j \left( \alpha \KH^{ij} \right) - h^{jk} \partial_j
  \partial_k \beta^i - \frac{1}{3} h^{ij} \partial_j \partial_k \beta^k,
\end{equation}
where
\begin{equation}
  \label{eq:hdef}
  h_{ij} = \gammaT_{ij} - \delta_{ij} \quad \mbox{and} \quad h^{ij} =
  \gammaT^{ij} - \delta^{ij}
\end{equation}
We call this condition as {\it the pseudo-minimal distortion
condition}.

As for the slicing condition, we choose the maximal slicing, $K = 0$. From
Eq.(\ref{eq:evoltrK}), we have an elliptic equation for the lapse
function $\alpha$:
\begin{equation}
  \label{eq:alph}
   D^m D_m \alpha = \alpha \left( \KH_{ij} \KH^{ij} + 4 \pi
   \left(\rhoh + S \right) \right) .
\end{equation}

\section{Numerical Methods}
As for the spatial coordinates in numerical calculations, we use a
Cartesian coordinate system $(x,y,z)$. To obtain initial data, we
should solve elliptic partial differential equations of
Eqs.(\ref{eq:mconst3}) and (\ref{eq:hconst2}).  The coupled elliptic
equations (\ref{eq:mconst3}) can be reduced to four decoupled Poisson
equations:
\begin{equation}
  \label{eq:spot2}
  \triangle \chi = 6 \pi \partial_i \JT_i ,
\end{equation}
\begin{equation}
  \label{eq:vpot2}
  \triangle W_i = 8 \pi \JT_i - \frac{1}{3} \partial_i \chi ,
\end{equation}
where $\chi = \partial_i W_i$.  These Poisson equations are solved
using a pre-conditioned conjugate gradient method.~\cite{ONS97} The
boundary conditions for $\chi$, $W_i$ and $\phi$ are, respectively,
given by
\begin{equation}
  \label{eq:chi}
  \chi = \frac{P_{i}x^{i}}{2r^3} - \frac{3M_{ij}}{2r^3} +
  \frac{9M_{ij}x^{i}x^{j}}{2r^5} + \mbox{O}\left(\frac{1}{r^4}\right),
\end{equation}
\begin{equation}
  \label{eq:Wi}
  W_{i} = -\frac{P_{k}x^{k}x^{i}}{4r^3} - \frac{7P_{i}}{4r}
   -\frac{(7M_{ij}-M_{ji}-M_{kk}\delta_{ij})}{4r^3}
  -\frac{3M_{jk}x^{j}x^{k}x^{i}}{4r^5}
  +\mbox{O}\left(\frac{1}{r^3}\right),
\end{equation}
and
\begin{equation}
  \phi = \frac{M}{r} + \frac{d_{k}x^{k}}{r^3} +
  \mbox{O}\left(\frac{1}{r^3}\right),
\end{equation}
for large $r$, where
\begin{equation}
  P_{i} = \int{\JT_{i}} \, \mbox{d}^3x, \qquad M_{ij} =
  \int{\JT_{i}x^{j}} \, \mbox{d}^3x ,
\end{equation}
\begin{equation}
  M= \int{S} \, \mbox{d}^3x, \qquad d_{k}=\int{Sx^k} \, \mbox{d}^3x
\end{equation}
and $S$ is the right-hand side of Eq.(\ref{eq:hconst2}).  Since the
source term $S$ depends on $\phi$, we will find a self-consistent
solution of Eq.(\ref{eq:hconst2}) iteratively. The non-linear
iteration will usually converge within 10 rounds.

At each time step, we should also solve the elliptic equation
(\ref{eq:beta}) for the shift vector $\beta^i$. The same procedure as
for $W_i$ is followed for it. Since the source term depends on
$\beta^i$, a self-consistent solution should be found iteratively. The
iteration will usually converge soon, but it sometimes becomes more
than 30 rounds at the final stage of a simulation.

Hydrodynamics equations (\ref{eq:hydrob}), (\ref{eq:hydrom}) and
(\ref{eq:hydroe2}) are solved using van Leer's scheme~\cite{VL23}.
The evolution equations of metric Eqs.(\ref{eq:evolphi}),
(\ref{eq:evolgTij}), (\ref{eq:evolkHij}) and (\ref{eq:evoltrK}) are,
respectively written as 
\begin{equation}
  \label{eq:evolphi1}
  \Dtv{- \beta}{\phi} = - \frac{\phi}{6}
  \left( \alpha K - \partial_l \beta^l \right) ,
\end{equation}
\begin{equation}
  \label{eq:evolgTij1}
  \Dtv{- \beta}{\gammaT_{ij}} = - 2 \left[ \alpha \KH_{ij}
    - \STF{\left( \gammaT_{il} \partial_j \beta^l \right)} \right],
\end{equation}
\begin{eqnarray}
  \Dtv{-\beta}{\KH_{ij}} & = &
  \phi^{-4} \left\{ \alpha \left[ \STF{(R_{ij})} - 8 \pi
        \STF{(S_{ij})} \right] - \STF{(D_i D_j \alpha)} \right\}
    \qquad \qquad \   \nonumber \\
    & + & \alpha \left( K \KH_{ij} - 2 \KH_{il} \KH^l{}_j \right)
    + \KH_{il} \partial_j \beta^l
    + \KH_{jl} \partial_i \beta^l
    - \frac{2}{3} \KH_{ij} \partial_l \beta^l
     \label{eq:evolkHij1}
\end{eqnarray}
and
\begin{equation}
  \Dtv{-\beta}{K}  =  \alpha \! \left[ \KH_{ij} \KH^{ij} \! + \!
    \frac{1}{3} K^2 
    \! + \! 4\pi \! \left( \rhoh \! + \! S^i{}_i \right) \right]
  \! - \! D^i D_i \alpha . \label{eq:evoltrK1}
\end{equation}
These equations are written as
\begin{equation}
  \label{eq:evmetric}
  \partial_t Q + v^\ell \partial_\ell Q = F ,
\end{equation}
with $v^\ell = - \beta^\ell$, which is
solved using the CIP (Cubic-Interpolated Pseudoparticle/Propagation)
method.\cite{YAB97} In the CIP method, both Eq.(\ref{eq:evmetric}) and
its spatial derivatives
\begin{equation}
  \label{eq:evmetricd}
  \partial_t (\partial_a Q) + v^\ell \partial_\ell (\partial_a Q) = -
  ( \partial_i v^\ell) (\partial_\ell Q) + \partial_a F
\end{equation}
are solved. This method reduces numerical diffusion during the
propagation of $Q$, since the time evolution of $Q$ and its
derivatives are traced.
Even if the gradient of $Q$ becomes very large near the surface of the
neutron star, the numerical diffusion is small so that errors and the
instability does not appear.

From the Hamiltonian constraint, the conformal factor $\phi$ also
obeys
\begin{equation}
  \widetilde{\Delta}\phi = -
  \frac{\phi^5}{8}\left(16\pi\rho_{H}+\KT_{ij}\KT^{ij} -
    \phi^{-4}\widetilde{R}\right).
  \label{eq:phi2}
\end{equation}
To obtain $\phi$, we first make $\phi$ evolve using
Eq.(\ref{eq:evolphi1}) and then calculate the right-hand side of
Eq.(\ref{eq:phi2}) using new $\phi$ and finally solve
Eq.(\ref{eq:phi2}).

For the evolution of both matter and metric, we use a two-step algorithm
to achieve second-order accuracy. That is, to solve the 
evolution equation such as
\begin{equation}
  \partial_t Q = F,
\end{equation}
first evolve half a time step $\frac{1}{2} \Delta t$,
\begin{equation}
  Q(t + \frac{1}{2}\Delta t) = Q(t) + \frac{1}{2} \Delta t \cdot F(t) .
\end{equation}
Then $F$ is calculated from quantities at $t + \frac{1}{2} \Delta
t$ and finally $Q(t + \Delta t)$ is calculated from $Q(t)$ and $F(t +
\frac{1}{2} \Delta t)$ as
\begin{equation}
  Q(t + \Delta t) = Q(t) + \Delta t \cdot F(t + \frac{1}{2} \Delta t) .
\end{equation}

We must take special care to calculate the Ricci tensor appearing on
the right-hand side of Eq.(\ref{eq:evolkHij1}). With the conformal
transformation of Eq. (\ref{eq:def-gammaTij}), the Ricci tensor
$R_{ij}$ associated with $\gamma_{ij}$ is given by
\begin{equation}
  R_{ij}=\RT_{ij} + R^{\phi}_{ij},
\end{equation}
where
\begin{equation}
  R^{\phi}_{ij} = -2 \phi^{-1} \left( \DT_{j}\DT_i \phi +
  \tilde{\gamma}_{ij}\widetilde{\Delta}\phi \right) +
  2 \phi^{-2}\left[ 3 (\DT_{i}\phi ) (\DT_{j}\phi ) -
    \tilde{\gamma}_{ij} ( \DT_{k}\phi ) ( \DT^{k}\phi ) \right]
\end{equation}
and $\RT_{ij}$ is the Ricci tensor associated with $\gammaT_{ij}$. The
tensor $\RT_{ij}$ is given by
\begin{equation}
  \label{eq:RT1}
  \RT_{ij} = \frac{1}{2}
  \left[\gammaT^{kl}(\gammaT_{li,jk}+\gammaT_{lj,ik} -
  \gammaT_{ij,kl}) + \gammaT^{kl}{}_{,l} (\gammaT_{ki,j} +
  \gammaT_{kj,i} - \gammaT_{ij,k}) \right] - \GammaT^{k}_{li}
  \GammaT^{l}_{jk},
\end{equation}
where $\GammaT^{i}_{jk}$ is the Christoffel symbol associated with
$\gammaT_{ij}$.  The second derivatives of $\gammaT_{ij}$ are replaced
by finite differences in numerical calculation but the numerical
precision of terms such as $\gammaT_{ij,kl}$ with $k{\neq}l$ is not so
good, while the degree of precision of $\gammaT_{ij,kl}$ with $k=l$ is
the same as that of the first derivatives.  Inaccuracies in
$\gammaT_{ij,kl}$ will cause a numerical instability. Since
\begin{equation}
  \label{eq:Gammadi}
  \gammaT^{jk} \gammaT_{ij,k} = - \gammaT_{ij} \gammaT^{jk}{}_{,k} =
  -\gammaT_{ij} \FT^j \equiv - \FT_i
\end{equation}
and
\begin{equation}
  \gammaT^{kl} \gammaT_{li,jk} = \FT_{i,j} - \gammaT^{kl}{}_{,j}
  \gammaT_{il,k} ,
\end{equation}
then
\begin{eqnarray}
  \RT_{ij} & = & \frac{1}{2} \left( \FT_{i,j} + \FT_{j,i} -
    \gammaT^{kl}{}_{,j} \gammaT_{il,k} - \gammaT^{kl}{}_{,i}
    \gammaT_{jl,k} - \gammaT^{kl} \gammaT_{ij,kl} \right) \nonumber \\
  & + & \frac{1}{2} \left[ \FT^k (\gammaT_{ki,j} + \gammaT_{kj,i} -
    \gammaT_{ij,k}) \right] - \GammaT^{k}_{li} \GammaT^{l}_{jk}.
  \label{eq:RT2}
\end{eqnarray}
On the {\it pseudo-minimal distortion condition}, where $\FT^i = \FT_i
= 0$, we have
\begin{equation}
  \RT_{ij} = - \frac{1}{2} \left( \gammaT^{kl}{}_{,j} \gammaT_{il,k} -
  \gammaT^{kl}{}_{,i} \gammaT_{jl,k} - \gammaT^{kl} \gammaT_{ij,kl}
  \right) - \GammaT^{k}_{li} \GammaT^{l}_{jk}.
  \label{eq:RT3}
\end{equation}
Although the second derivative appears in the term $\gammaT^{kl}
\gammaT_{ij,kl}$, it can be written as
\begin{equation}
  \gammaT^{kl} \gammaT_{ij,kl} = (\delta^{kl} + h^{kl}) h_{ij,kl} =
  h_{ij,kk} + h^{kl} h_{ij,kl} ,
\end{equation}
and thus inaccuracies in numerical value of $\gammaT_{ij,kl}$ will not
be so serious, if both $h_{kl}$ and $h_{ij,kl}$ are small.

Boundary conditions for $\KH_{ij}$ and $\gammaT_{ij}$ are important,
because the boundary is not so far from the stars and therefore
inappropriate boundary conditions cause reflections of the outgoing
waves. We apply outgoing conditions for $\KT_{ij}$ and $\gammaT_{ij}$,
in our coordinate conditions. The condition can be written as
\begin{equation}
Q(t,x^i)=\frac{H({\alpha}t-{\phi}^{2}r)}{r},
\end{equation}
where $r=\sqrt{x^{2}+y^{2}+z^{2}}$. Here H satisfies an advection
equation
\begin{equation}
  \partial_{t}H+c^{l}\partial_{l}H=0.
\end{equation}
where
\begin{equation}
  c^{l}=\frac{{\alpha}x^l}{r{\phi}^2}.
\end{equation}
This implies
\begin{equation}
  \partial_{t}Q+c^{l}{\partial}_{l}Q=-\frac{\alpha}{r{\phi}^{2}}Q.
  \label{325}
\end{equation}
Eq.(\ref{325}) is solved along with Eqs.(\ref{eq:evolgTij1}) and
(\ref{eq:evolkHij1}) using the CIP method.

\section{Gauge-Invariant Wave Form Extraction}

In the gauge conditions described in \S\ref{sec:CC}, non-wave parts of
the perturbations decrease as $O(r^{-1})$ for large $r$ and therefore
$h_{ij}$ defined by Eq.(\ref{eq:hdef}) cannot be simply considered as
the gravitational waves~\cite{ONS97}; it includes gauge dependent
modes. So that, it is necessary to perform a gauge-invariant wave form
extraction. Here we apply a gauge-invariant wave form extraction
technique suggested by Abrahams et.al.~\cite{ADHS92}, which is based
on Moncrief's formalism~\cite{MON74}.

Outside of the star, we can consider the total spacetime metric
$g_{\mu\nu}$ which is generated numerically as a sum of a
Schwarzschild spacetime and non-spherical perturbation parts:
\begin{equation}
  g_{\mu\nu}=g^{(B)}_{\mu\nu}+h_{\mu\nu}^{(e)}+h_{\mu\nu}^{(o)},
\end{equation}
where $g^{(B)}_{\mu\nu}$ is a spherically symmetric metric given by
\begin{equation}
  g^{(B)}_{\mu\nu}= 
  \left(\begin{array}{cccc} -N^2 & 0 & 0 & 0 \\ 0 &
      A^2 & 0 & 0 \\ 0 & 0 & R^2 & 0 \\ 0 & 0 & 0 & R^2\sin^2\theta
    \end{array}\right)
\end{equation}
and $h^{(e)}_{\mu\nu}$ and $h^{(o)}_{\mu\nu}$ are even-parity and
odd-parity metric perturbations, respectively. They are given by
\begin{equation}
  h^{(e)}_{\mu\nu} = \sum_{lm} \left(\begin{array}{@{}c@{\ }c@{\
  }cc@{}} N^2H_{0lm}Y_{lm} & H_{1lm}Y_{lm} &
  h^{(e)}_{0lm}Y_{lm,\theta} & h^{(e)}_{0lm}Y_{lm,\phi} \medskip \\
  \mbox{sym} & A^2H_{2lm}Y_{lm} & h^{(e)}_{1lm}Y_{lm,\theta} &
  h^{(e)}_{1lm}Y_{lm,\phi} \medskip \\ \mbox{sym} & \mbox{sym} &
  \displaystyle r^2 \left[
  K_{lm}+G_{lm}\frac{\partial^2}{\partial\theta^2} \right] Y_{lm} &
  r^2G_{lm}X_{lm} \medskip \\ \mbox{sym} & \mbox{sym} & \mbox{sym} &
  h^{(e)}_{33}
    \end{array}
  \right)
\label{eq:heven}
\end{equation}
and
\begin{equation}
  h^{(o)}_{\mu\nu} = \sum_{lm}\left(
    \begin{array}{cccc}
      0 & 0 & \displaystyle
      -h^{(o)}_{0lm}\frac{Y_{lm,\phi}}{\sin\theta} &
      h^{(o)}_{0lm}Y_{lm,\theta} \sin{\theta} \medskip \\ 0 & 0 &
      -h^{(o)}_{1lm}Y_{lm,\phi}\sin{\theta} &
      h^{(o)}_{1lm}Y_{lm,\theta}\sin{\theta} \medskip \\ \mbox{sym} &
      \mbox{sym} & \displaystyle
      \frac{1}{2}h^{(o)}_{2lm}\frac{X_{lm}}{\sin\theta} &
      \displaystyle -\frac{1}{2}h^{(o)}_{2lm}W_{lm}\sin{\theta}
      \medskip \\ \mbox{sym} & \mbox{sym} & \mbox{sym} & \displaystyle
      -\frac{1}{2}h^{(o)}_{2lm}X_{lm} \sin{\theta}
    \end{array}
  \right),
\label{eq:hodd}
\end{equation}
where
\begin{equation}
  h^{(e)}_{33} = r^2\sin^2\theta \left[ K_{lm}Y_{lm} + G_{lm} \left(
  \frac{\partial^2}{\partial\theta^2} Y_{lm} - W_{lm} \right) \right],
\end{equation}
$N^2$, $A^2$, $R^2$, $H_{1lm}$, $h^{(e)}_{0lm}$, $h^{(e)}_{1lm}$,
$K_{lm}$, $G_{km}$, $h^{(o)}_{0lm}$, $h^{(o)}_{1lm}$, and
$h^{(o)}_{2lm}$ are the functions of $t$ and $r$ and $Y_{lm}$ is the
spherical harmonics.  The functions $X_{lm}$ and $W_{lm}$ are
given by
\begin{equation}
  X_{lm} = 2\left(
    \frac{\partial^2}{\partial{\phi}\partial{\theta}} -\cot\theta
    \frac{\partial}{\partial\phi} \right) Y_{lm},
\end{equation}
and
\begin{equation}
  W_{lm} = \left(
    \frac{\partial^2}{\partial\theta^2} -\cot\theta
    \frac{\partial}{\partial\theta} -\frac{1}{\sin^2\theta}
    \frac{\partial^2}{\partial\phi^2} \right) Y_{lm},
\end{equation}
respectively.
The symbol `sym' in Eqs.(\ref{eq:heven}) and (\ref{eq:hodd}) indicates
the symmetric components.  From the linearized theory about
perturbations of the Schwarzschild spacetime, the gauge invariant
quantities $\Psi^{(o)}$ and $\Psi^{(e)}$ are given by
\begin{equation}
  \Psi^{(o)}_{lm}(t,r) = \sqrt{2\Lambda(\Lambda-2)} N^2 \frac{1}{r}
  \left( h^{(o)}_{1lm} + \frac{r^2}{2} \frac{\partial}{{\partial}r}
  \left( \frac{h^{(o)}_{2lm}}{r^2} \right) \right)
\end{equation}
and
\begin{equation}
  \Psi^{(e)}_{lm}(t,r) = -\sqrt{\frac{2(\Lambda-2)}{\Lambda}}
  \frac{4rN^2k_{2lm} + \Lambda rk_{1lm}}{(\Lambda + 1 - 3N^2)}
\end{equation}
for the odd and even parity modes, respectively, where
\begin{equation}
  \Lambda = l(l+1),
\end{equation}
\begin{equation}
  k_{1lm} = K_{lm} + N^2 rG_{lm,r} - 2\frac{N^2}{r} h^{(o)}_{1lm}
\end{equation}
and
\begin{equation}
  k_{2lm} = \frac{H_{2lm}}{2N^2} - \frac{1}{\sqrt{N^2}}
  \frac{\partial}{{\partial}r} \left( \frac{r}{N^2}K_{lm} \right).
\end{equation}
The quantities $\Psi^{(o)}$ and $\Psi^{(e)}$ satisfy the Regge-Wheeler
and the Zerilli equations, respectively,
\begin{equation}
  \left[ \frac{\partial^2}{{\partial}t^2} -
  \frac{\partial^2}{{\partial}r_{\ast}^2} + V^{(I)}\right]
  \Psi^{(I)}_{lm}=0, \quad (I=\mbox{e, o}),
\end{equation}
where $V^{(o)}$ and $V^{(e)}$ are the Regge-Wheeler and the Zerilli
potentials\cite{ZER24}. The symbol $r_{\ast}$ is the tortoise
coordinate defined by $r_\ast = r + 2M \ln (r/2M -1)$.  Two
independent polarizations of gravitational waves $h_{+}$ and
$h_{\times}$ are given by
\begin{equation}
  h_{+} - \mbox{i} h_{\times} = \frac{1}{\sqrt{2}r} \sum_{l,m}
  (\Psi^{(e)}_{lm}(t,r) + \Psi^{(o)}_{lm}(t,r))_{-2}Y_{lm},
\end{equation}
where
\begin{equation}
  {}_{-2}Y_{lm} = \frac{1}{\sqrt{\Lambda(\Lambda-2)}} \left( W_{lm} -
  \frac{\mbox{i}}{\sin \theta} X_{lm} \right).
\end{equation}
In numerical calculations, the functions $N^2(t,r)$, $A^2(t,r)$ and
$R^2(t,r)$ of the background metric are calculated by performing the
following integration over a two-sphere of radius $r$:
\begin{eqnarray}
  N^2(t,r) &=& -\frac{1}{4\pi} \int g_{tt} \, \dOmega ,
  \label{eq:N2} \\ 
  A^2(t,r) &=& \frac{1}{4\pi} \int g_{rr} \, \dOmega ,
  \label{eq:A2} \\
  R^2(t,r) &=& \frac{1}{8\pi} \int \left( g_{\theta\theta} +
    \frac{g_{\phi\phi}}{\sin^2\theta} \right) \, \dOmega,
  \label{eq:R2}
\end{eqnarray}
where $\dOmega = \sin \theta \mbox{d}\theta \mbox{d}\phi$.  The
components of the even parity metric perturbations are
\begin{equation}
  H_{2lm}(t,r) = \frac{1}{A^2} \int g_{rr} Y^{\ast}_{lm} \, \dOmega ,
\end{equation}
\begin{equation}
  G_{lm}(t,r) = \frac{1}{\Lambda(\Lambda-2)} \frac{1}{R^2} \int \left[
  \left( g_{\theta\theta} - \frac{g_{\phi\phi}}{\sin^2\theta} \right)
  W^{\ast}_{lm} + \frac{2g_{\theta\phi}}{\sin\theta} \,
  \frac{X^{\ast}_{lm}}{\sin\theta} \right] \, \dOmega,
\end{equation}
\begin{equation}
  K_{lm}(t,r) = \frac{1}{2}\Lambda G_{lm} +\frac{1}{2R^2} \int \left(
  g_{\theta\theta} +\frac{g_{\phi\phi}}{\sin^2\theta} \right)
  Y^{\ast}_{lm}) \, \dOmega 
\end{equation}
and
\begin{equation}
  h^{(e)}_{1lm}(t,r) = \frac{1}{\Lambda} \int \left(
  g_{r\theta}Y^{\ast}_{lm,\theta} + \frac{g_{r\phi}}{\sin\theta} \,
  \frac{Y^{\ast}_{lm,\phi}}{\sin\theta} \right) \, \dOmega,
\end{equation}
where $\ast$ denotes the complex conjugate.  The components of odd
parity metric perturbations are
\begin{equation}
  h^{(o)}_{1lm}(t,r) = -\frac{1}{\Lambda} \int \left( g_{r\theta}
  \frac{Y^{\ast}_{lm,\phi}}{\sin\theta} -\frac{g_{r\phi}}{\sin\theta}
  Y^{\ast}_{lm,\theta} \right) \, \dOmega 
\end{equation}
and
\begin{equation}
   h^{(o)}_{2lm}(t,r) = \frac{1}{\Lambda(\Lambda-2)} \int \left[
   \left( g_{\theta\theta} - \frac{g_{\phi\phi}}{\sin^2\theta} \right)
   \frac{X^{\ast}_{lm}}{\sin\theta} - \frac{2g_{\theta\phi}}{\sin\phi}
   W^{\ast}_{lm} \right] \, \dOmega.
\label{eq:ho2}
\end{equation}

The metric tensor $g_{rr}$, $g_{\theta\theta}$ etc.~on the spherical
coordinate system appearing in Eqs.(\ref{eq:A2})--(\ref{eq:ho2}) are
calculated from $g_{xx}$, $g_{yy}$ etc.~on the Cartesian coordinate
system, such as
\begin{eqnarray}
  g_{rr} & = & \Sts (g_{xx} \Cps + 2 g_{xy} \Sp\Cp + g_{yy} \Sps)
  \nonumber \\ & & + g_{zz} \Cts + 2\St\Ct ( g_{yz}\Sp + g_{zx} \Cp).
\end{eqnarray}

We need angular integrals over spheres for constant $r$, such as
\begin{equation}
  \label{eq:IntdOmega}
  F(r_0) = \int_{r = r_0} f(x,y,z) \,\dOmega = \int f(r_0,\theta,\phi)
  \,\dOmega,
\end{equation}
while we have the physical quantities only at Cartesian grid points.
We need interpolation to obtain the values of $f(r_0,\theta,\phi)$ from
$f(x,y,z)$ at the grid points. It is, however, not easy to fully
parallelize the procedure on a parallel computer with distributed
memory.  We therefore rewrite Eq.(\ref{eq:IntdOmega}) as a volume
integral, namely,
\begin{equation}
  \label{eq:IntdV}
  F(r_0) = \int f(x,y,z) \delta(r-r_0) \, \mbox{d}^3 x =
  \lim_{a\rightarrow 0} \frac{1}{\sqrt{\pi} a r_0^2} \int F(x,y,z)
  e^{-(r-r_0)^2/a^2} \, \mbox{d}^3 x ,
\end{equation}
where $r = \sqrt{x^2 + y^2 + z^2}$. Numerical integral with $a =
\Delta x/2$ gives a good value to Eq.(\ref{eq:IntdV}), where $\Delta
x$ is the separation between grid points. It is favorable both in the
accuracy and the speed for parallel computers.

\section{Numerical results}

\begin{figure}[tbp]
  \centerline{\includegraphics[width=\textwidth]{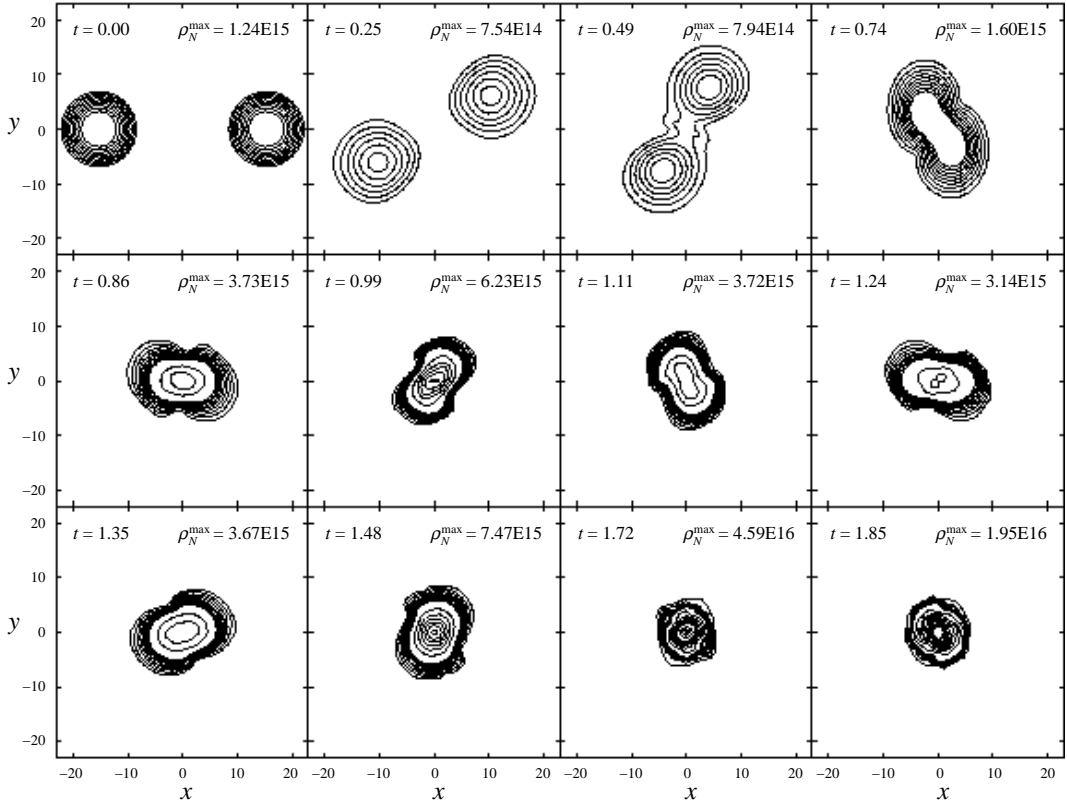}}
  \caption{Density $\rhon$ on the $x$-$y$ plane. Time $t$ in units of
    milliseconds and the maximum of density $\rhon$ at each time are
    shown. Contour lines are drawn $a \times 10^{b}$g/cm$^{3}$, where
    $a = 1, 2, \cdots 9$ and $b$ = 14, 15, 16. The unit of the length
    is $1\sol$.}
  \label{fig:denxy}
\end{figure}
We performed numerical simulations for coalescing binary neutron stars
and evaluated the gravitational waves. We used $475\times 475\times
238$ Cartesian grid assuming the symmetry with respect to the
equatorial plane, which require memory of about 80GBytes. The
separation between grid points is $\Delta x = \Delta y = \Delta z =
1\sol$, i.e., equal spacing.  Here we use the units of $G = c = 1$.
As the initial condition, we put two spherical stars of rest mass
$1.5\sol$ and radius $7.7 \sol = 11.6$km. The separation between the
center of each mass is $30.2 \sol = 46.2$km.  As for an equation of
state, we use the ${\gamma}=2$ polytropic equation of state. The
initial rotational velocity is given so that the circulation of the
system vanishes approximately as
\begin{equation}
  \vec{V}_a(\vec{r}) = \vec{\Omega} \times \vec{r} - \vec{\Omega}
  \times \left( \vec{r} - \vec{r}_a \right),
  \qquad (a = 1, 2)
\end{equation}
where $\vec{\Omega}$ is the orbital angular velocity and $\vec{r}_a$
is the location of the center of each star.  Now we set
$\Omega = 0.010/\sol$, then the angular momentum $J_0$ becomes
$6.7\sol^2$. The total ADM mass $M_{\rscript{ADM}}$ of the system is
$2.76\sol$ and $q \equiv J_0/M_{\rscript{ADM}}^2 = 0.89$.
\begin{figure}[tbp]
  \begin{minipage}[t]{.49\textwidth}
  \centerline{\includegraphics[width=.95\textwidth]{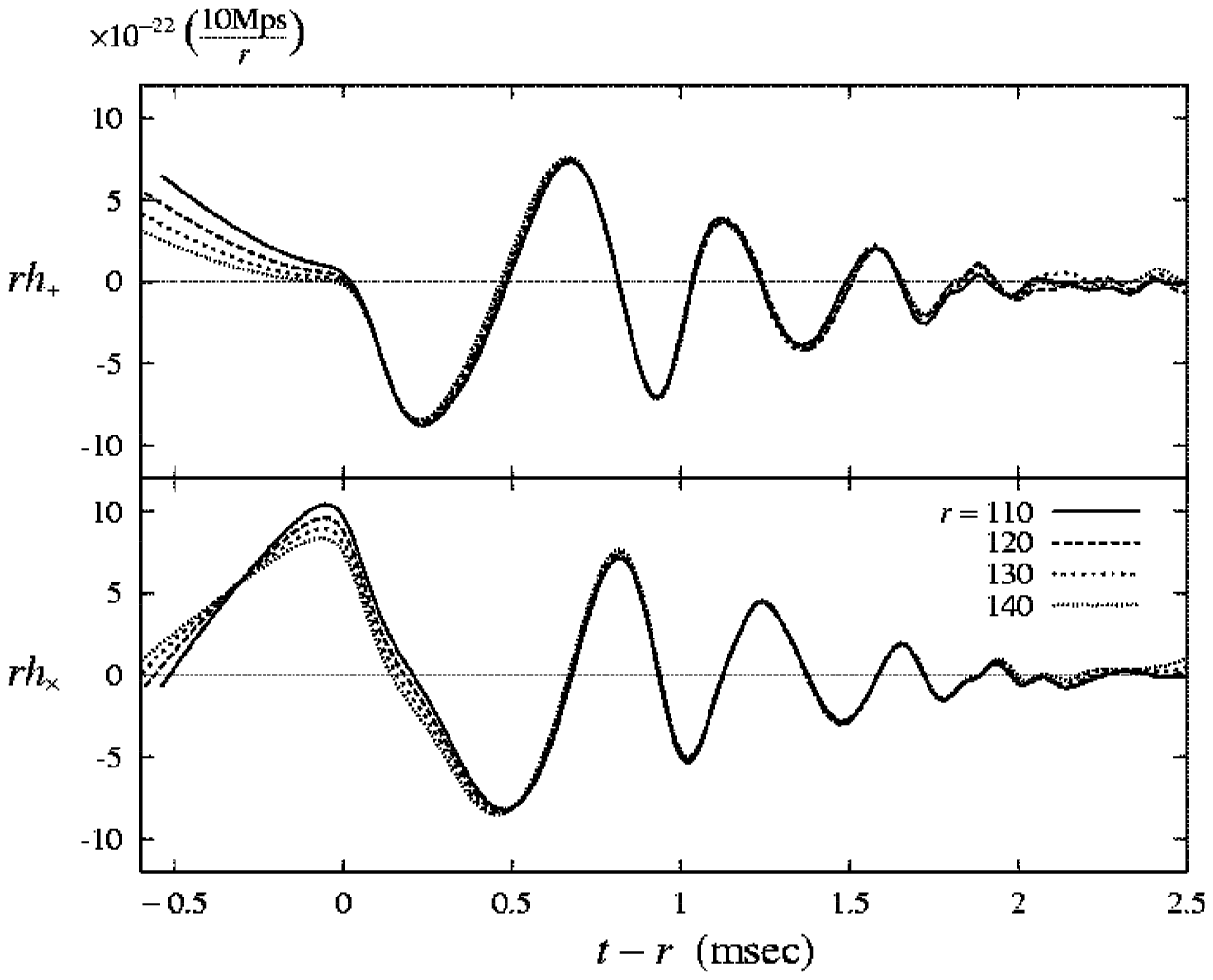}}
  \caption{Plots $r h_{+,\times}$ along $z$-axis at $r = 110 \sim
    140 \sol$ as a function of $t-r$.}
    \label{fig:wave0}
  \end{minipage}
  \hfill
  \begin{minipage}[t]{.49\textwidth}
  \centerline{\includegraphics[width=.95\textwidth]{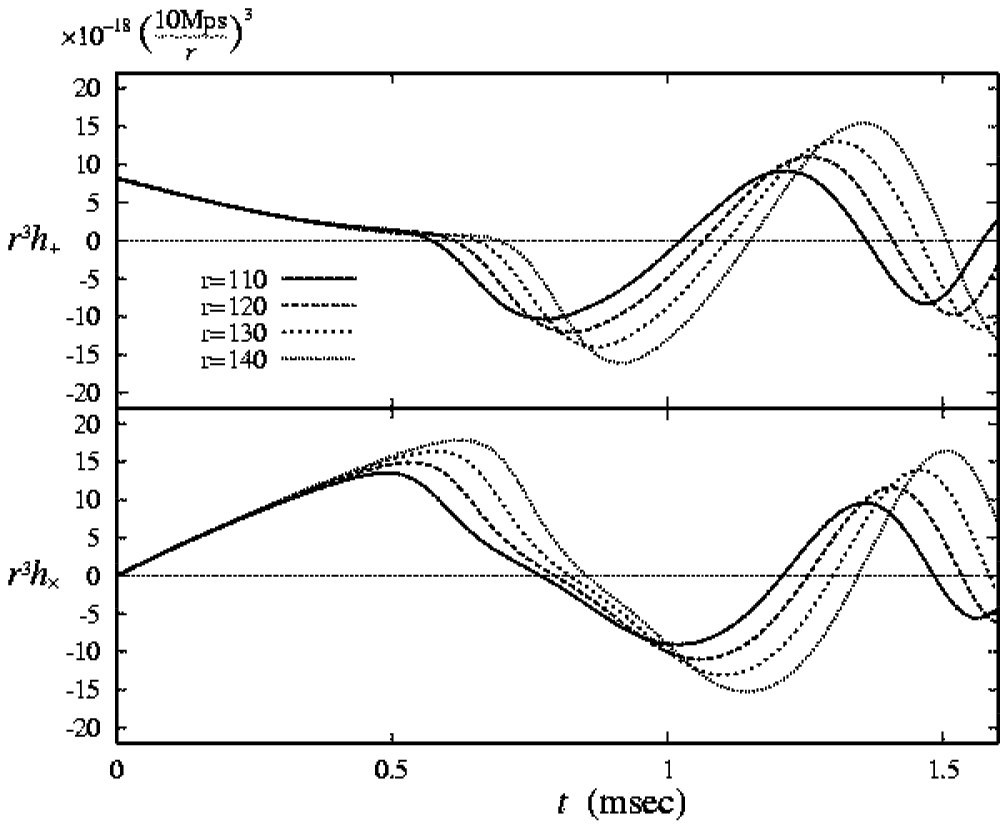}}
  \caption{Plots $r^3 h_{+,\times}$ as a function of $t$.}
    \label{fig:wave3}
  \end{minipage}
\end{figure}

Figures \ref{fig:denxy} shows the evolution of the density $\rhon$ on
the $x$-$y$ plane. The stars start to coalesce at approximately $t =
0.5$msec and an almost axisymmetric star is formed by $t=1.8$msec.

Figure \ref{fig:wave0} shows the gravitational wave forms $rh_+$ and
$rh_\times$ on the $z$-axis evaluated at $r=110, 120, 130$ and $140
\sol$ as functions of the retarded time $t-r$.  The lines of
$rh_{+}(t-r)$ and $rh_{\times}(t-r)$ estimated at $r = 110 \sim 140
\sol$ for $t-r > 0$ coincide with each other but they don't for $t-r <
0$. Figure \ref{fig:wave3} shows $r^3 h$ as a function of $t$. It
reveals that $h$ includes a non-wave mode proportional to $r^{-3}$,
which corresponds to the quadrupole part in the Newtonian potential of
the background metric. It decreases fast as the merger of stars
proceeds.

Therefore we eliminate the non-wave mode from $h_{+}$ and $h_{\times}$
using Fourier transformation as follows:
\begin{itemize}
\item Assuming that total waves are expressed as a sum of wave parts
$F(t-r)/r$ and non-wave parts $G(t)/r^3$,
  \begin{equation}
    h(t,r)=\frac{F(t-r)}{r}+\frac{G(t)}{r^3}.
  \end{equation}
\item Fourier components of $h(t,r)$ are written as
  \begin{equation}
    h_{\omega}(r) = \frac{e^{-\rscript{i}{\omega}r}}{r}F_{\omega}(r) +
    \frac{1}{r^3}G_{\omega}(r).
  \end{equation}
  where
  \begin{equation}
    F_{\omega} \equiv \frac{1}{2\pi}
    \int F(t) e^{-\rscript{i}\omega t} \,\mbox{d}t
  \end{equation}
  and
  \begin{equation}
    G_{\omega} \equiv \frac{1}{2\pi}
    \int G(t) e^{-\rscript{i}\omega t} \,\mbox{d}t.
  \end{equation}
\item From the values of $h_{\omega}(r)$ in different radial
coordinates $r_1$ and $r_2$, $F_{\omega}$ can be given by
  \begin{equation}
    F_{\omega}=\frac{r_{2}^{3}h_{\omega}(r_2)-r_{1}^{3}h_{\omega}(r_1)}
    {r_{2}^{2} e^{-\rscript{i} \omega r_2}
      - r_{1}^{2} e^{-\rscript{i} \omega r_1}}.
  \end{equation}
\item Fourier components $\hat{h}_{\omega}(r)$ of gravitational waves
are given by
  \begin{equation}
    \hat{h}_{\omega}(r_k) \equiv
    \frac{e^{-\rscript{i} \omega r_k}}{r_k}F_{\omega}, \quad k=1,~2.
  \end{equation}
\item By inverse Fourier transformation of $\hat{h}_{\omega}(r)$, we
can get the gravitational waves, which do not include non-wave modes,
  \begin{equation}
    \label{eq:defhpx}
    h_{+}(t,r) - \mbox{i} h_{\times}(t,r) =\int{\hat{h}_{\omega}(r)
    e^{\rscript{i} \omega t}}\,\mbox{d}\omega.
  \end{equation}
\end{itemize}
\begin{figure}[tbp]
  \begin{minipage}{.49\textwidth}
  \centerline{\includegraphics[width=\textwidth]{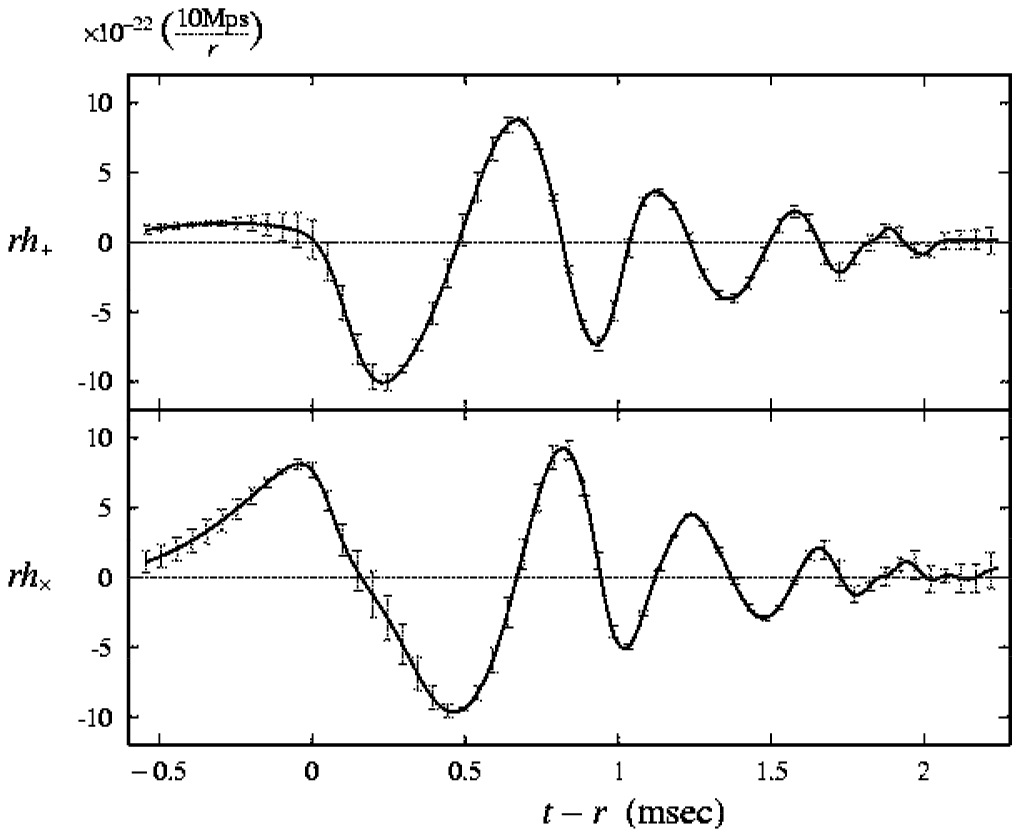}}
  \caption{Wave forms $r h_{+,\times}$ along $z$-axis as a function of
    $t-r$. The curves are averages of $r h_{+,\times}$ estimated at $r
    = 110 \sim 200 \sol$ and error bars denote $2\sigma$.}
    \label{fig:wave}
  \end{minipage}
  \hfill
  \begin{minipage}{.45\textwidth}
  \centerline{\includegraphics[width=\textwidth]{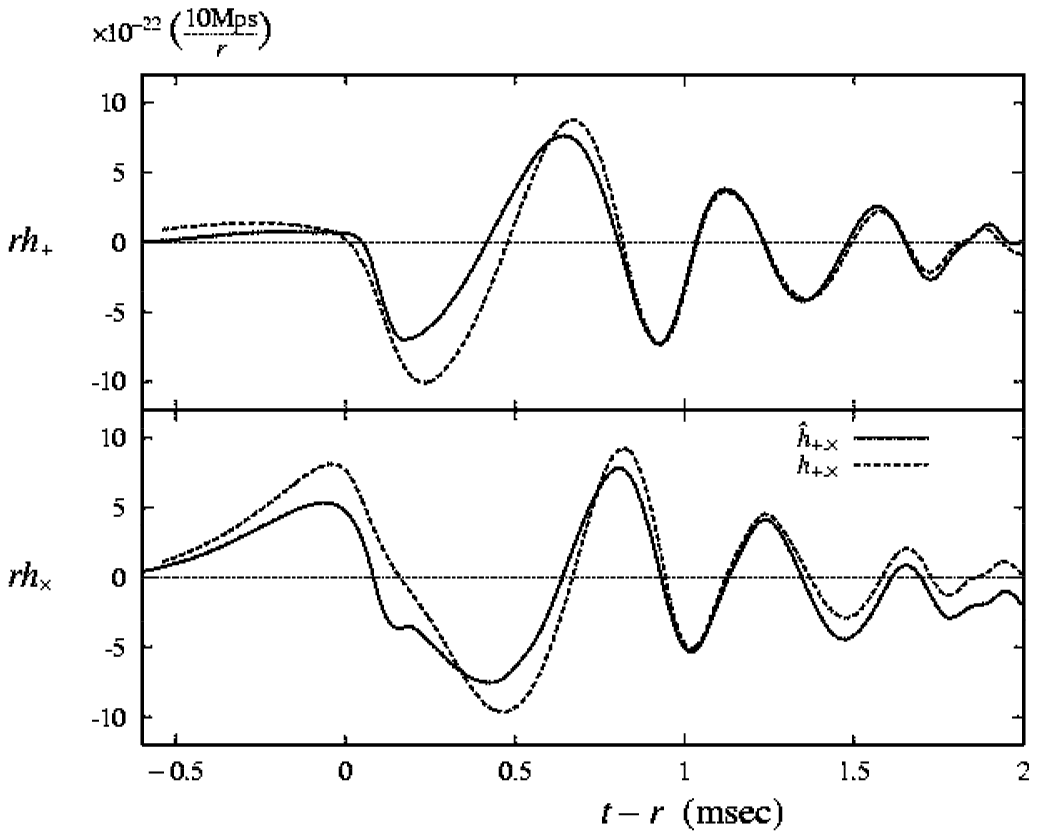}}
  \caption{The comparison $h_{+,\times}$ shown in Fig.\ref{fig:wave}
    (solid lines)and $\widehat{h}_{+,\times}$ defined by
    Eqs.(\ref{eq:hp}) and (\ref{eq:hx}) (dashed lines).}
    \label{fig:wavec}
  \end{minipage}
\end{figure}
\begin{figure}[tbp]
  \centerline{\includegraphics[width=.9\textwidth]{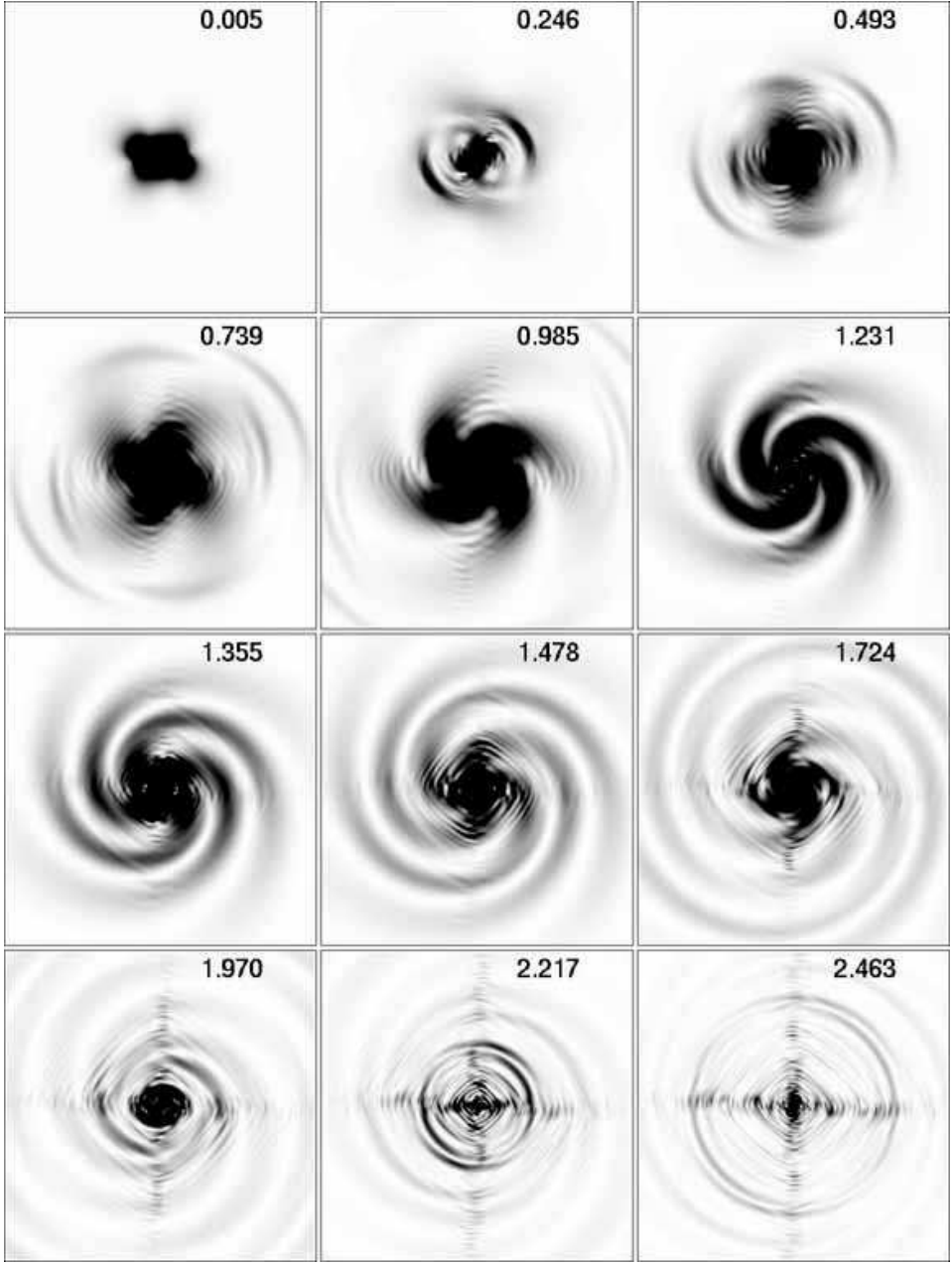}}
  \caption{The propagation of the gravitational waves. ``The energy
    densities of gravitational waves'' $r^2 \rho_{_{\rscript{GW}}}$ on
    the $x$-$y$ plane are shown as gray scale figures.  Time in units
    of milliseconds is shown.}
  \label{fig:gwxy}
\end{figure}
The resultant wave form is shown in Fig.\ref{fig:wave}. The curves
represent the average of $h_+$ and $h_\times$ calculated at $r = 110,
120, \cdots 200\sol$ and twice the dispersion $2\sigma$ is shown as
error bars.

Figure \ref{fig:wavec} compares $h_{+}$ and $h_\times$ given by
Eq.(\ref{eq:defhpx}) with the metric perturbation $\widehat{h}_+$ and
$\widehat{h}_\times$ defined by
\begin{equation}
  \label{eq:hp}
  \widehat{h}_+ = \frac{1}{2} \left( h_{xx} - h_{yy} \right),
\end{equation}
and
\begin{equation}
  \label{eq:hx}
  \widehat{h}_\times  =  h_{xy},
\end{equation}
respectively,
where $h_{ij}$ is defined by Eq.(\ref{eq:hdef}).  Although
$\widehat{h}_+$ and $\widehat{h}_\times$ include gauge dependent mode,
they almost coincide with the gauge independent wave $h_+$ and
$h_\times$. It shows that the gauge mode in $\widehat{h}_+$ and
$\widehat{h}_\times$ is small, while it may be only
accidental. However, we note that the pseudo-minimal distortion
condition Eq.(\ref{eq:pmind}) guarantee $\sum_{j} \partial_j h_{ij} =
0$, and thus $\widehat{h}_+$ and $\widehat{h}_\times$ are
transverse-traceless.  Then we can argue ``the energy density of the
gravitational waves'' as
\begin{equation}
  \label{eq:rhogw}
  \rho_{\rscript{GW}} = \frac{1}{32\pi} \dot{\gammaT}_{ij}
  \dot{\gammaT}_{ij} = \frac{1}{32\pi} \AT_{ij} \AT_{ij} .
\end{equation}
The propagation of $r^2 \rho_{\rscript{GW}}$ is shown in
Fig.\ref{fig:gwxy}. A spiral pattern appears, which can be explained
naively by the quadrupole wave pattern given by
\begin{equation}
  \label{eq:quadru}
  r^2 \rho_{\rscript{GW}} = \frac{r^2}{32\pi} \left( \AT_{ij}
  \right)^2 \propto \cos2 \theta + \sin^2 \theta \, \sin^2 \left(
  2\Omega (t-r) - 2 \phi\right)/4.
\end{equation}
On the $x$-$y$ plane, where $\theta = \pi/2$, $\rho_{\rscript{GW}}$ is
constant along the spiral of $r + \phi/\Omega$ = constant.

\begin{figure}[tbp]
  \centerline{\includegraphics[width=.6\textwidth]{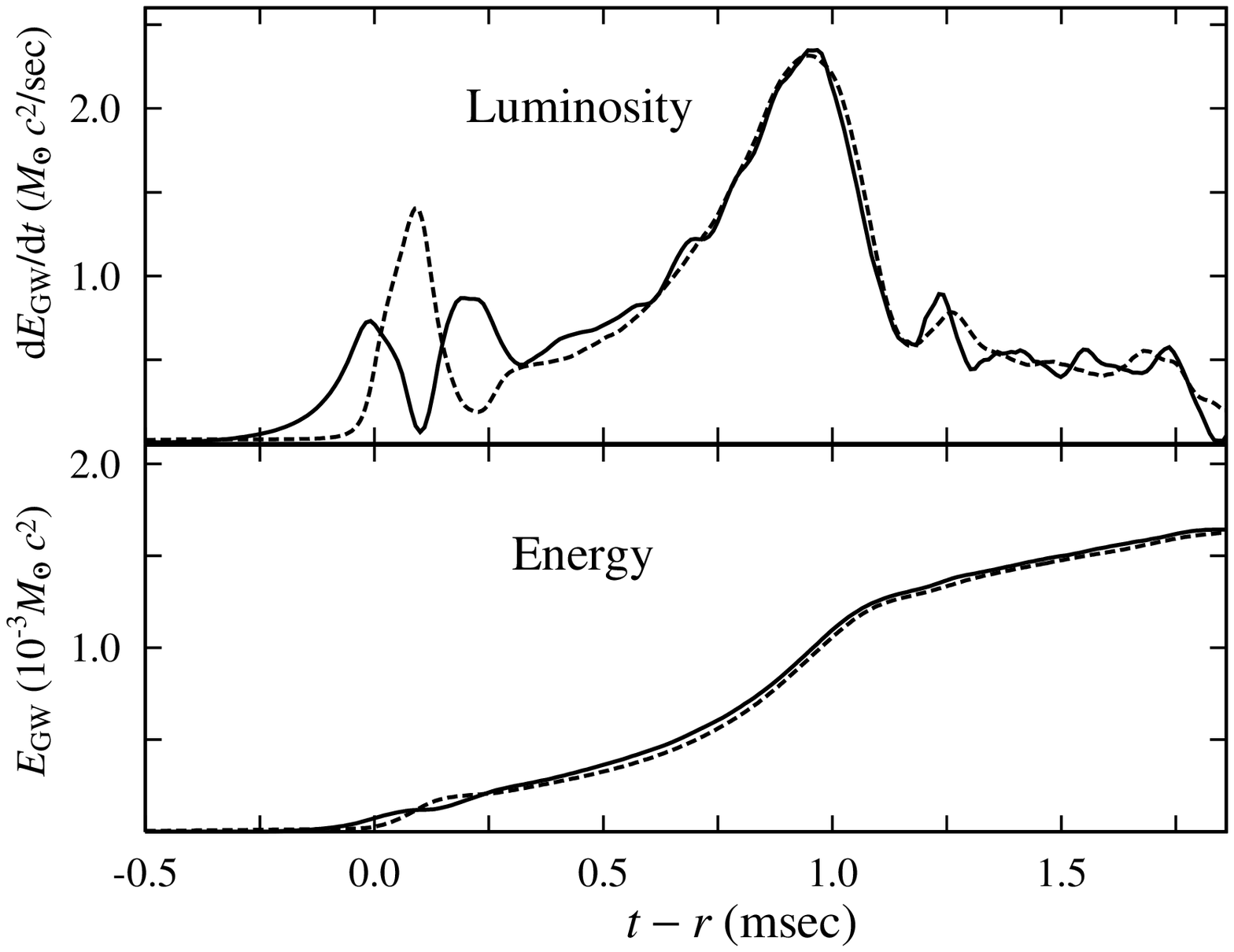}}
  \caption{The luminosity $\mbox{d}E_{\rscript{GW}}/\mbox{d}t$
    (in units of $\sol c^2$/sec) and total energy $\Delta
    E_{\rscript{GW}}$ (in units of $10^{-3} \sol c^2$) emitted as
    gravitational waves up to $t$. The solid and dashed lines plot the
    values calculated using Eq.(\ref{eq:dEdt}) and Eq(\ref{eq:dEdtg}),
    respectively. They are evaluated at $r = 200\sol$.}
    \label{fig:flux}
\end{figure}

The gauge invariant luminosity $\mbox{d}E_{\rscript{GW}}/\mbox{d}t$
and the total energy $\Delta E_{\rscript{GW}}$ of the gravitational
waves can be calculated as
\begin{equation}
  \label{eq:dEdt}
  \frac{\mbox{d}E_{\rscript{GW}}}{\mbox{d} t} = \frac{1}{32\pi}
  \sum_{l,m} \left( \left| \partial_t \Psi^{(e)}_{lm}(t,r) \right|^2 +
  \left| \partial_t \Psi^{(o)}_{lm}(t,r) \right|^2 \right)
\end{equation}
and
\begin{equation}
  \label{eq:Egw}
  \Delta E_{\rscript{GW}} = \int
  \frac{\mbox{d}E_{\rscript{GW}}}{\mbox{d} t} \,\mbox{d}t ,
\end{equation}
respectively.
The luminosity and the total energy can be also estimated from
Eq.(\ref{eq:rhogw}) as
\begin{equation}
  \label{eq:dEdtg}
  \frac{\mbox{d} \widetilde{E}_{\rscript{GW}}}{\mbox{d} t} = \int_{r
  \rightarrow \infty} \, r^2 \rho_{\rscript{GW}} \,\dOmega
\end{equation}
and
\begin{equation}
  \label{eq:Egwg}
  \Delta \widetilde{E}_{\rscript{GW}} = \int \frac{\mbox{d}
  \widetilde{E}_{\rscript{GW}}}{\mbox{d} t} \,\mbox{d}t,
\end{equation}
which is not gauge invariant. The luminosity and the total energy
emitted as gravitational waves up to $t$ calculated at $r = 200\sol$
using Eqs.(\ref{eq:dEdt}) and (\ref{eq:Egw}) as well as
Eqs.(\ref{eq:dEdtg}) and (\ref{eq:Egwg}) are plotted as 
functions of $t-r$ in Fig.\ref{fig:flux}, which shows that
Eq.(\ref{eq:rhogw}) gives a good estimate of the energy density of the
gravitational waves as expected from Fig.\ref{fig:wavec}.

\section{Conclusion}
We showed a stable code using {\it the pseudo-minimal distortion
condition} and the maximal slicing condition for a coalescing neutron
star binary.

\begin{figure}[tbp]
  \centerline{\includegraphics[width=.8\textwidth]{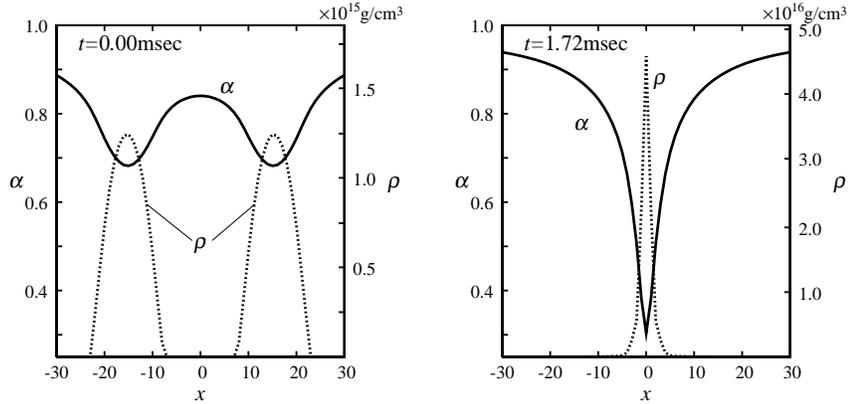}}
  \caption{Lapse function $\alpha$ (solid lines) and density $\rhon$
    (dashed lines) along $x$-axis at $t$=0(left) and 1.72msec(right).}
    \label{fig:alphrho}
\end{figure}
We were able to extract the wave form of the gravitational radiation
using the gauge-invariant extraction techniques.  The amplitude of the
gravitational waves is
\begin{equation}
  h \sim 1 \times 10^{-21} \left(\frac{r}{\mbox{10Mpc}}\right).
\end{equation}
As shown in Fig.\ref{fig:flux}, the total energy emitted is
\begin{equation}
  \Delta E_{\rscript{GW}} (t < 3\mbox{msec}) \approx 2 \times 10^{-3}
  \sol c^2 \approx 3 \times 10^{51} \mbox{erg} \approx 0.1\% \mbox{ of
  } M_{\rscript{tot}} \, c^2 .
\end{equation}
The angular momentum lost by the gravitational waves are given by
\begin{equation}
  \label{eq:DeltaJ}
  \Delta J = \frac{1}{32\pi} \int \sum_{l,m} \left( \left| m
  \Psi^{(e)}_{lm} \, \partial_t \Psi^{(e)}_{lm} \right| + \left| m
  \Psi^{(o)}_{lm} \, \partial_t \Psi^{(o)}_{lm} \right| \right)
  \,\mbox{d} t,
\end{equation}
which is $7 \times 10^{-2} \, (G \sol^2/c) \sim $ 1\% of the initial
angular momentum of the system.

We have not searched an apparent horizon to see if a black hole is
formed, Instead we plot $\alpha$ and $\rhon$ along the $x$-axis at
$t$=0 and 1.72msec in Fig.\ref{fig:alphrho}. The gradient of $\alpha$
near the surface of the merged star becomes large and the value gets
less than 0.3. This suggests the formation of a black hole.

If the black hole is formed after the merger of two neutron stars,
quasi-normal modes of the black hole are excited. To investigate a
possibility that the excitation of the quasi-normal modes can be seen
by the numerically calculated waves, we evaluated the energy spectrum
of the gravitational waves, which is given by
\begin{equation}
  \frac{\mbox{d}E_{\rscript{GW}}}{\mbox{d}\omega}= \frac{1}{32\pi}
  \sum_{l,m} \omega^2 \left( \left| \Psi^{(e)}_{lm\omega}(r) \right|^2
  + \left| \Psi^{(o)}_{lm\omega}(r) \right|^2 \right),
\end{equation}
\begin{wrapfigure}{r}{\halftext}
  \centerline{\includegraphics[width=.37\textwidth]{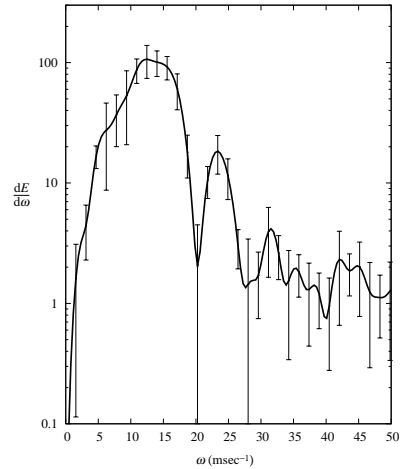}}
  \caption{The energy spectrum of the gravitational waves.
    The curves are averages of $\mbox{d}E/\mbox{d}\omega$ estimated at
    $r = 110 \sim 200 \sol$ and error bars denote $2\sigma$.}
    \label{fig:spec}
\end{wrapfigure}
where $\Psi^{(I)}_{lm\omega}(r)$ is the Fourier transformation of
$\Psi^{(I)}_{lm}(t,r)$, and is shown in Fig.~\ref{fig:spec}.  The
frequencies of the quasi-normal modes depend on the angular momentum
of the black hole, but the fundamental frequency of $l = 2$ for a
Schwarzschild black hole of mass $2.8\sol$ is
${\omega}=25~\mbox{msec}^{-1}$. A peak near this frequency appears in
Fig.~\ref{fig:spec}.  Unfortunately, however, rotating angular
frequency just when the merger of the stars finishes is
$12~{\sim}~15~\mbox{msec}^{-1}$ and thus they will radiate the waves
of frequency near ${\omega}=25~\mbox{msec}^{-1}$.  So it is not clear
whether this peak corresponds to the emission of the quasi-normal mode
of the formed black hole.

Our results must be compared with the model H-1 of Shibata and
Uryu.\cite{SU02}. The wave form and the luminosity of the
gravitational waves are qualitatively consistent, while our values of
the radiated energy and angular momentum are smaller than theirs. It
caused by the difference of the initial data; two stars merge more
quickly in our calculation since they are not in quasi-equilibrium at
the initial time. Nevertheless, the ratio of $(\Delta
E_{\rscript{GW}}/M_{\rscript{ADM}})$ to $(\Delta J/J_0)$ agrees with
their discussion.

Finally, we must mention the accuracy of our code.  Essential parts of
code tests are summarized in Oohara and Nakamura.\cite{ONS97} We need
not solve Eq.(\ref{eq:evoltrK}) for $K$ since we use the maximal
slicing, where $K =0$, but it is solved to monitor the numerical
precision. The value of $K$ is kept as small as 0.1\% of a typical
value of $K_{ij}$ but it becomes large from the numerical boundary
region in the final stage of simulation. It is due to a small
reflection of gravitational waves at the numerical boundary. Since we
use the pseudo-minimal distortion condition, $\FT^{i} =
\gammaT^{ij}{}_{,j}$ must be zero. It is satisfied within a few \% of
the derivative of $\gammaT^{ij}$, while the error grows in the same
way as $K$ in final stage of simulation. The total ADM mass is
conserved within 10\% error up to $t < 1.7$msec. When two stars merge
and a single compact object is formed, the conservation get worse,
since our grid is too coarse to represent such a compact object. Since
it is likely to be a black hole by that time, however, the global
characteristics of gravitational radiation should be hardly affected.

We found that the gravitational waves are in small part reflected at the
numerical boundary and reflected waves grow up gradually after $t
\approx 2.5$msec. It is because the boundary is still too close. In
reality, if the distance of the numerical boundary is halved, the growth
of the reflected waves near the boundary will be visible by $t =
1.4$msec in the distribution of the energy density of the
gravitational waves $\rho_{_{\rscript{GW}}}$, while it is not yet
visible in Fig.\ref{fig:gwxy}.  This problem must be overcome if the
numerical boundary is located farther. In addition, a finer grid is
required to investigate the formation of a black hole.  To perform
simulations with a finer and larger-sized grid, we must improve
schemes for solving elliptic partial differential equations and
evolution equations.  The most CPU hours are consumed for solving
elliptic equations for $\beta^i$ (Eqs.(\ref{eq:chi}) and
(\ref{eq:Wi})), $\alpha$ (Eq.(\ref{eq:alph})) and $\phi$
(Eq.(\ref{eq:phi2})), and thus the reduction of CPU hours required to
solve these equations is effective. The present code requires memory
of 80GBytes and 100 CPU hours with a $475\times 475\times 238$ grid.
The size is not restricted by the memory but by CPU hours. Therefore,
larger-scale simulations can be performed according to the speedup of
the code. We are still improving the code, including implementation of
new schemes and an apparent horizon finder as well as reformulation of
the Einstein equations, some results of which will be presented soon.

\section*{Acknowledgment}

Numerical computations were carried out on SR8000/F1 at Hight Energy
Accelerator Research Organization(KEK) and on VPP5000 at the
Astronomical Data Analysis Center of the National Astronomical
Observatory Japan(NAO).  This work was in part supported by
Grant-in-Aid for Scientific Research (C), No.13640271, from
Japan Society for the Promotion of Science, by the Supercomputer
Project No.097(FY2003) of KEK and by the Large Scale Simulation
Project yko12a of NAO. This work was also supported in part by
Grant-in-Aid for Scientific Research of the Japanese Ministry of
Education, Culture, Sports, Science and Technology, No.14047212 (TN)
and No.14204024 (TN).

%


\begin{thebibliography}{99}
\bibitem{ONS97} K.~Oohara, T.~Nakamura and M.~Shibata,
\PTPS{128,1997,183}.
\bibitem{ON99} K.~Oohara and T.~Nakamura, \PTPS{136,1999,270}
\bibitem{DETECT} Refer Web sites;
http://tamago.mtk.nao.ac.jp/tama.html, http://www.ligo.caltech.edu/,
http://www.pg.infn.it/virgo/ and http://www.geo600.uni-hannover.de/.
\bibitem{HPA91} P.~Haensel, B.~Paczynski and P.~Amsterdamski,
\AJ{375,1991,209}.
\bibitem{RM92}M.~Rees and P.~Meszaros,
\JL{Mon.~Not.~R.~Astron.~Soc.,258,1992,41}.
\bibitem{ADHS92} A.~Abrahams, D.~Dernstein, D.~Hobill and E.~Seidel,
\PRD{45,1992,3544}.
\bibitem{SU02} M.~Shibata and K.~Uryu, \PTP{107,2002,265}.
\bibitem{SY03} For example, H.~Shinkai and G.~Yoneda in {\em Progress
in Astronomy and Astrophysics} (Nova Science Publ., 2003) and
references therein.
\bibitem{SN95} M.~Shibata and T.~Nakamura, \PRD{42,1995,5428}.
\bibitem{BS99} T.W.~Baumgarte and S.L.~Shapiro, \PRD{59,1999,064037}.
\bibitem{NOK87} T.~Nakamura, K.~Oohara and Y.~Kojima,
\PTPS{90,1987,1}.
\bibitem{ON97} K.~Oohara and T.~Nakamura, in {\em Relativistic
Gravitation and Gravitational Radiation} (Cambridge Univ. Press,
1997), p.~309.
\bibitem{NO99} T.~Nakamura and K.~Oohara, in {\em Numerical
Astrophysics}, (Kluwer Academic Publ., 1999), p.~247.
\bibitem{KO02} K.~Oohara and T.~Nakamura, in {\em Proceedings of the
Ninth Marcel Grossmann Meeting on General Relativity} (World
Scientific, 2002), p.~2295.
\bibitem{MON74} V.~Moncrief, \JL{Ann.~Phys.~(N.Y.),88,1974,322}.
\bibitem{ZER24} F.~J.~Zerilli, \PRL{24,1970,737}.
\bibitem{VL23} B.~van Leer, \JL{J.~Comput.~Phys.,23,1997,276}.
\bibitem{YAB97} T.~Yabe, in {\em Computational Fluid Dynamics Review
1997} (Wiley, 1997).

\end{thebibliography}
\end{document}